\documentstyle[amssymb,aps,prl,epsfig]{revtex} 

\begin{document}
\draft

\title{Anomalies in the conduction edge of quantum wires}
\author{T. Rejec$^1$, A. Ram{\v s}ak$^{1,2}$ and J.H. Jefferson$^3$}
\address{$^1$J. Stefan Institute, SI-1000 Ljubljana, Slovenia\\
$^2$Faculty of Mathematics and Physics, University of Ljubljana,
SI-1000 Ljubljana, Slovenia\\
$^3$DERA, St. Andrews Road, Great Malvern, Worcestershire WR14 3PS,
England}
\date{July 26, 2000}
\maketitle

\begin{abstract}
\widetext
\smallskip
We study the conductance threshold of
clean nearly straight quantum wires in which an
electron is bound. We show that such a system exhibits spin-dependent
conductance structures on the rising
edge to the first conductance plateau, one near $0.25(2e^2/h)$,
related to a singlet resonance, and one near
$0.75(2e^2/h)$, related to a triplet
resonance. As a quantitative example we solve
exactly the scattering problem for two-electrons in a wire with
circular cross-section and a weak bulge. From the scattering matrix we
determine conductance via the Landauer-B\" uttiker
formalism.  The conductance
anomalies are robust and survive to temperatures of a few degrees.
With increasing magnetic field the conductance exhibits a plateau at
$e^2/h$, consistent with recent experiments.
\end{abstract}

\section{Introduction}
Following the pioneering work in Refs. \cite{wees88,wharam88} many
groups have now observed conductance steps in
various types of quantum wire. These first experiments were performed
on gated two-dimensional electron gas (2DEG) structures, while similar
behaviour of conductance are shown in ``hard-confined'' quantum wire
structures, produced by cleaved edge over-growth \cite{yacoby96},
epitaxial growth on ridges \cite{ramvall97}, heteroepitaxial growth on
``v''-groove surfaces \cite{walther92} and most recently in
GaAs/Al$_{\delta}$Ga$_{1-\delta}$As narrow ``v''-groove
\cite{kaufman99} and low-disorder \cite{kane98} quantum wires.

These experiments strongly support the idea of ballistic conductance in
quantum wires and are in surprising agreement with the now standard
Landauer-B\"{u}ttiker formalism \cite{landauer57,buttiker86} neglecting
electron interactions \cite{houten92}. However, there are certain anomalies,
some of which are believed to be related to electron-electron interactions
and appear to be spin-dependent. In particular, already in early experiments
a structure is seen in the rising edge of the conductance curve
\cite{wees88}, starting at around $0.7 G_0$ with $G_0=2e^{2}/h$ and merging
with the first conductance plateau with increasing energy. Under increasing
in-plane magnetic field, the structure moves down, eventually merging with a
new conductance plateau at $e^{2}/h$ in very high fields
\cite{thomas96,thomas98}. Theoretically this anomaly has not been adequately
explained, despite several scenarios, including spin-polarised sub-bands
\cite{fasol94}, conductance suppression in a Luttinger liquid with repulsive
interaction and disorder \cite{maslov95} or local spin-polarised
density-functional theory \cite{wang98}. Recently we have shown that these
conductance anomalies near $0.7G_0$ and $0.25G_0$ are consistent with an
electron being weakly bound in wires of circular and rectangular
cross-section, giving rise to spin-dependent scattering resonances
\cite{rejec99,rejec00,flambaum99}.

In this paper we develop further the single bound-electron picture and give
new results for wires of circular cross-section, including magnetic field
dependence.

\section{The model}
\begin{figure}[htb]
\center{\epsfig{file=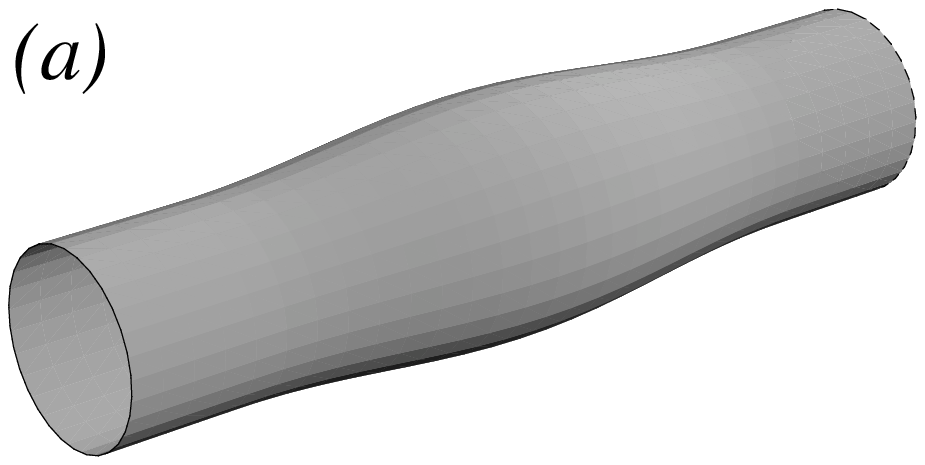,height=30mm,angle=-0}
\epsfig{file=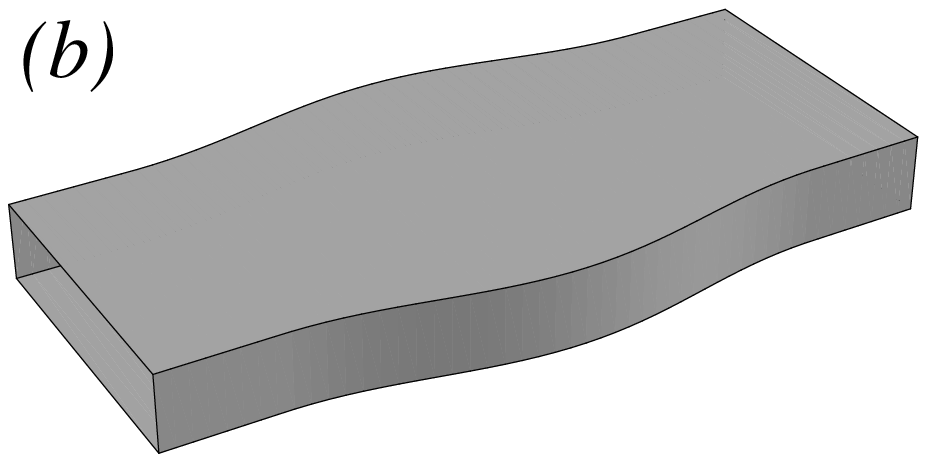,height=30mm,angle=-0}}
\caption{The wire shape is symmetric around the
$z$-axis. The potential is con\-stant $V(x,y,z)=0$ within the boundary,
and $V_{0}>0$ elsewhere.
(a) Circular cross-section, defined by,
$r_{0}(z)=\frac{1}{2}a_{0}( 1+\protect\xi \cos^{2}\protect\pi
z/a_{1})$   for $|z|\leq\frac{1}{2}a_{1}$ and $x_{0}(z)\equiv
\frac{1}{2}a_{0}$ otherwise.
(b)
Rectangular cross-section is defined by $x_{0}(z)=\frac{1}{2}a_{0}(1+\xi
\cos ^{2}\pi z/a_{1})$. }
\label{fig1}
\end{figure}

We consider quantum wires which are almost perfect but for which there is a
very weak effective potential, which has at most two bound states. Such an
effective potential can arise, for example, from a smooth potential due to
remote gates. Alternatively it could arise from a slight buldge in the an
otherwise perfect wire. We consider this latter situation for the cases of
quantum wires with both circular cross-section \cite{rejec99}, appropriate for
'hard-confined' v-groove wires; or rectangular cross-section
\cite{rejec00}, which
approximate 'soft-confined' wires resulting from gated 2DEGs.  These are
shown schematically in Figure~\ref{fig1}.
The cross-sections of these wires are
sufficiently small that the lowest transverse channel approximation is
adequate for the energy and temperature range of interest. The smooth
variation in cross-section also guarantees that inter-channel mixing is
negligible. Restricting ourselves to this lowest transverse channel, the
Schr\"{o}dinger equation on a finite-difference grid in the $z$-direction
may be written,
\begin{eqnarray}
H=-t\sum_{i\sigma }\left( c_{i+1,\sigma }^{\dagger }c_{i\sigma
}+c_{i\sigma }^{\dagger }c_{i+1\sigma }\right) +\sum_{i\sigma}\epsilon
_{i\sigma}n_{i\sigma}+
 \label{hubbard}
 \sum_{i}U_{ii}n_{i\uparrow }n_{i\downarrow }+\frac{1}{2}\sum_{i\neq
j}U_{ij}n_{i}n_{j},  
\end{eqnarray}
\noindent where $c_{i\sigma }^{\dagger }$ creates an electron with spin $\sigma $ at
the $z=z_{i}$ in the lowest transverse channel; $n_{i}=\sum_{\sigma
}n_{i\sigma }$ with $n_{i\sigma }=c_{i\sigma }^{\dagger }c_{i\sigma }$; $%
t=\hbar ^{2}/(2m^{\ast }\Delta ^{2})$, where $\Delta =z_{i+1}-z_{i};$ $%
\epsilon _{i}=\hbar ^{2}/(m^{\ast }\Delta ^{2})+\epsilon \left(
z_{i}+\right) + g^*\mu_B \sigma B $, where $\epsilon \left( z_{i}\right) $
is the energy of the lowest transverse channel at $z_{i}$ and $ g^*\mu_B
\sigma B $ is the Zeeman energy for a magnetic field $B$ in the
$z$-direction, as in Refs.~\cite{rejec99,rejec00}.
$U_{ij}$ is an effective screened Coulomb interaction which
was obtained by starting with a full 3D Coulomb interaction, integrating
over the lowest transverse modes and then adding screening
phenomenologically. The dielectric constant is taken as $\varepsilon=12.5$,
appropriate for GaAs. Note that this is a general form, the difference
between the two cases shown in Figure~\ref{fig1}
being reflected entirely in the
energy parameters $\epsilon$ and $U$. We note that this Hamiltonian also has
the form for a perfectly straight wire subject to a smooth potential
variation, defined by the $\epsilon$.

\section{Two-electron approximation}

Over a range of parameters in which the deviation from a perfect straight
wire is small, one, and only one, electron resides in a bound state. This is
because the weak effective potential provided by the bulge will always
contain at least one bound state and since the binding energy is small, a
second electron cannot be bound due to Coulomb repulsion. For larger bulges
or deeper potential wells more electrons may be bound but these situations
will not be considered here. Note that even if there is more than one bound
state (we consider one or two) there can still only be one bound electron
due to the Coulomb repulsion. The actual number of electrons in the wire may
be changed by varying the Fermi energies in the leads and reservoirs to
which the wire is connected. In experiments, this is achieved simply by
changing the voltages on one or more gate electrodes. When there is more
than one electron in the wire, a current will flow from source to drain
contacts and at low temperatures the motion of these electrons is ballistic.
The conduction electrons will scatter from the bound electron giving rise to
resistance. If we neglect the mutual interaction between conduction
electrons, then the transport problem reduces to a two-electron scattering
problem described by equation (\ref{hubbard}). This is a reasonable
approximation provided that the mean electron density is not too low, i.e.
that the mean electron separation is of order the effective Bohr radius or
less.  We solve the two-electron scattering problem exactly subject to the
boundary condition that the asymptotic states consist of one bound electron
in the ground state and one free electron. The main features of these
scattering solutions may be understood by the following simple physical
picture. The effective potential due to the single bound electron and the
effective potential well gives rise to a symmetric double barrier structure
since an incident electron will initially feel the Coulomb repulsion due to
the bound electron but will then pass through a weak local minimum, e.g., at
the the point of maximum diameter in the wire of circular cross-section.
This gives rise to a resonance, the peak of which corresponds to perfect
transmission. A more refined analysis shows that this resonance is spin
dependent, singlet and triplet resonances occuring at different energies,
with the singlet always lowest. In fact this spin-dependent scattering is
the quasi one-dimensional analogue of the three-dimensional case, discussed
some 70 years ago by Oppenheimer and Mott \cite{oppenheimer28}.

From the scattering solutions we compute the conductance using again a
Landauer-B\"{u}ttiker formula which, incorporating the results of
spin-dependent scattering, takes the following form,
\begin{eqnarray}
G=-\frac{2e^2}{h}\int
\frac{\partial f(\epsilon -E,T)}{\partial \epsilon
} 
 \left[\frac{1}{4}[{\cal T}_{\rm s}(\epsilon-E_B)+{\cal T}_{\rm
t}(\epsilon-E_B)]+
\frac{1}{2}{\cal T}_{{\rm t}}(\epsilon+E_B)\right]
{\rm
d}\epsilon \label{conductance}
\end{eqnarray}
where the subscripts ${\cal T}_{\rm s}$ and ${\cal T} _{\rm t}$ refer to
singlet and triplet transmission probabilities respectively, with
$E_B=\frac{1}{2} g^*\mu_B B$ and $E$ is the Fermi energy in the leads.
\section{Results}
We have performed detailed calculations for both the circular and
rectangular wire cross sections shown in Figure~\ref{fig1}. Apart from small
quantitative differences, the results are very similar and hence, for
brevity, we shall only show results for wires with circular cross-section.

\begin{figure}[htb]
\center{\epsfig{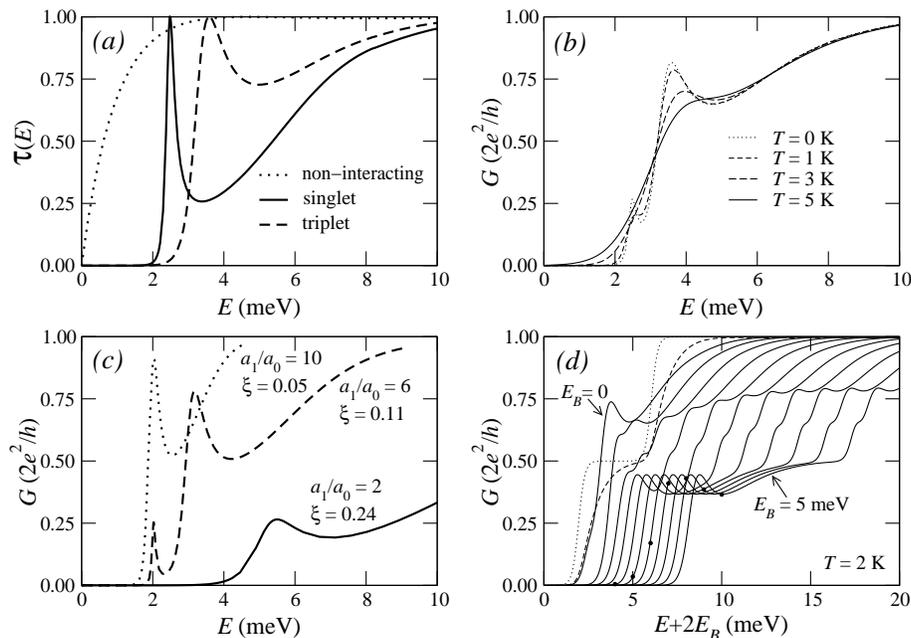}}
\caption{Resonance and conductance curves. (a) Typical transmission
probabilities for singlet and triplet resonances in zero magnetic
field. Here [and in (b) and (d)] $a_0=10$~nm, $a_1=60$~nm, $\xi=0.1$
and $V_0=0.4$~eV. (b)
Total conductance in zero magnetic field showing temperature dependence of
anomalies near 0.25 and 0.7. (c) Dependence of conductance on wire geometry.
(d) Magnetic field dependence of conductance showing weak resonance and
saturation to $e^2/h$ in a high magnetic field. }
\label{fig2}
\end{figure}

In Fig.~\ref{fig2}(a) we show plots at zero temperature and magnetic
field of ${\cal T}_{{\rm s}}(E)$ and ${\cal T}_{t}(E)$ for a typical
wire with the geometry of Fig.~\ref{fig1}(a).  The thin dotted line
represents the non-interacting result, independent of spin. We see
clearly the sharp singlet resonance at low energy followed by the
broader triplet at higher energy. In Fig.~\ref{fig2}(b) the
conductance $G$ in units of $2e^2/h$ is shown, as calculated for
various temperatures. The resonances have a strong temperature
dependence and, in
particular, the sharper singlet resonance is more readily washed out
at finite temperatures. However, it should be noted that resonances
survive to relatively high temperatures, because the width of the
wire, which dictates the energy scale, is small ($a_0=10$~nm)
\cite{ramsak98}.  Note that for weak coupling, the energy scale is set
by the $x$-energy of the lowest channel, $\sim a_{0}^{-2}$ and hence
the conductance vs. $Ea_{0}^{2}$ with $U a_0$ fixed is roughly
independent of $a_{0}$ (the scaling would be exact for
$V_0\to\infty$).

In Fig.~\ref{fig2}(c) we show the zero temperature and field conductance
curves for three different bulge shapes, which become longer and flatter as
we move from right to left. For the case with the shortest bulge region
(right) we see only a singlet resonance. This is because the effective
potential well has only one bound state and hence, even in the absence of
Coulomb interaction, would not support a triplet. On the other hand, the
other two cases have two one-electron bound states. In the absence of
Coulomb interaction, these levels give rise to singlet and triplet bound
states. When the Coulomb interaction is switched on, the states develop into
two resonance peaks (dotted line). Here the position of both peaks nearly
coincides, because the buldge is relatively long and the singlet-triplet
splitting is small. For the remaining case (dashed line), both the lowest
singlet and triplet develop into resonances when the Coulomb interaction is
switched on, though the singlet is only just unbound with the resonance
lying close to the conduction edge energy and is thus very sharp.

In Fig.~\ref{fig2}(d) conductance for $T=2$~K is presented for magnetic field increasing
from zero in steps with $\Delta E_B=0.5$~meV and for clarity the
curves have been shifted by $2E_B$ to the right with increasing
$E_B$. We present results for $a_0=10$~nm, but note that $E_B$ also
obeys the above mentioned scaling $E_B a_0^2$ with varying
$a_0$. Magnetic fields which would give substantial effects in e.g.
narrow ``v''-groove wires \cite{kaufman99}, would have to be very
large, since $E_B=1$~meV corresponds to large $g^*B\sim
35$~T. However, due to ``$E_B a_0^2$'' scaling, the corresponding
value for a wider wire with $a_0\sim50$~nm would be only $\sim 1.4$~T.
Also plotted in Fig.~\ref{fig2}(d) for comparison are the
corresponding results for the non-interacting electron case (dotted)
and the perfectly straight wire (dashed), with $E_B=2$~meV. In this
figure we have indicated with a dot the points $E=E_B$. To the left of
these points $G$ simplifies, $G(E,B)=\frac{e^2}{h}{\cal T}_{\rm
t}(E+E_B)$, whereas at high energies spin-flip transmission
probabilities $t_{\uparrow \downarrow \to \uparrow \downarrow}$ and
$t_{\uparrow \downarrow \to \downarrow \uparrow}$, contained within
the remaining terms of Eq.~(\ref{conductance}), are non-zero. These
parts of the curves should be treated with caution though they are
expected to be more reliable at lower fields.

\section{Summary}
In summary, we have shown that quantum wires with weak longitudinal
confinement, or open quantum dots, can give
rise to spin-dependent, Coulomb blockade
resonances when a single electron is bound in the confined region.
The emergence of a specific structure at $G(E)\sim \frac{1}{4}G_{0}$
and $G\sim \frac{3}{4}G_{0}$ is a consequence of the singlet and
triplet nature of the resonances and the probability ratio 1:3 for
singlet and triplet scattering and as such is a universal effect. A
comprehensive numerical investigation of open quantum dots using a
wide range of parameters shows that singlet resonances are always at
lower energies than the triplets, in accordance with the corresponding
theorem for bound states \cite{lieb62}.  With increasing in-plane
magnetic field, the resonances shift their position and a plateau
$G(E)\sim e^2/h$ emerges. The effect of a magnetic field is observable
only in relatively wider quantum wires, due to the intrinsic energy
scale $\propto a_0^{-2}$.

\begin{acknowledgments}
The authors wish to acknowledge A.V. Khaetskii and C.J. Lambert for helpful
comments. This work was part-funded by the U.K. Ministry of Defence and the
EU.
\end{acknowledgments}




\begin{references}
\bibitem{wees88}  {B.J. van Wees {\em et al.}, Phys. Rev. Lett. {\bf 60},
848 (1988).}
\bibitem{wharam88}  {D.A. Wharam {\em et al.}, J. Phys. C {\bf 21}, L209
(1988).}
\bibitem{yacoby96}  {A. Yacoby {\em et al.}, Phys. Rev. Lett. {\bf 77}, 4612
(1996).}
\bibitem{ramvall97}  {P. Ramvall {\em et al.}, Appl. Phys. Lett. {\bf 71},
918 (1997).}
\bibitem{walther92}  {M. Walther, E. Kapon, D.M. Hwang, E. Colas, and L.
Nunes, Phys. Rev. B {\bf 45}, 6333 (1992); M. Grundmann {\em et al.},
Semicond. Sci. Tech. {\bf 9}, 1939 (1994); R. Rinaldi {\em et al.},
Phys. Rev. Lett. {\bf 73}, 2899 (1994).}
\bibitem{kaufman99}  {D. Kaufman {\em et al.}, Phys. Rev. B {\bf 59}, R10433
(1999).}
\bibitem{kane98} {B.E. Kane {\em et al.}, Appl. Phys. Lett. {\bf 72},
3506 (1998); D.J. Reilly, cond-mat/0001174.}
\bibitem{landauer57}  {R. Landauer, IBM J. Res. Dev. {\bf 1}, 223 (1957);
{\bf 32}, 306 (1988).}
\bibitem{buttiker86}  {M. B\"{u}ttiker, Phys. Rev. Lett. {\bf 57},
1761 (1986).}
\bibitem{houten92} {H. van Houten, C.W.J. Beenakker, and B.J. van Wees, in
{\it Semiconductors and Semmimetals},  edited by R.K. Willardson,
A.C. Beer, and E.R. Weber, (Academic Press, 1992).}
\bibitem{thomas96}  {K.J. Thomas {\em et al.}, Phys. Rev. Lett. {\bf 77},
135 (1996); Phys. Rev. B {\bf 58}, 4846 (1998); {\bf 59}, 12252 (1999).}
\bibitem{thomas98} {K.J. Thomas {\em et al.}, Phil. Mag. B {\bf 77},
1213 (1998).}
\bibitem{fasol94}  {G. Fasol and H. Sakaki, Jpn. J. Appl.
Phys. {\bf 33}, 879 (1994).}
\bibitem{maslov95} {D.L. Maslov, Phys. Rev. B {\bf 52}, R14368,
1995}.
\bibitem{wang98}  {Chuan-Kui Wang and K.-F. Berggren, Phys. Rev. B {\bf 57},
4552 (1998).}
\bibitem{rejec99} {T. Rejec, A. Ram\v sak, and J.H. Jefferson,
cond-mat/9910399.}
\bibitem{rejec00} {T. Rejec, A. Ram\v sak, and J.H. Jefferson,
J. Phys.: Condens. Matter {\bf 12}, L233 (2000).}
\bibitem{flambaum99} {V.V. Flambaum and M.Yu. Kuchiev, cond-mat/9910415.}
\bibitem{oppenheimer28} {J.R. Oppenheimer, Phys. Rev. {\bf 32}, 361
(1928); N.F. Mott, Proc. Roy. Soc. A {\bf 126}, 259 (1930).}
\bibitem{ramsak98}  {A. Ram\v sak, T. Rejec, and J. H. Jefferson, Phys. Rev.
B {\bf 58}, 4014 (1998).}
\bibitem{lieb62} {E. Lieb and D. Mattis, Phys. Rev. {\bf 125}, 164
(1962).}
\end{references}
\end{document}